\documentclass[10pt,conference]{IEEEtran}

\usepackage{ifpdf}
\usepackage{cite}

\ifCLASSINFOpdf
\usepackage{graphicx}
\else
\fi

\usepackage[cmex10]{amsmath}
\usepackage{algorithmic}
\usepackage{fixltx2e}
\usepackage{stfloats}
\usepackage{url}

\begin{document}
\title{Cross-Layer Design to Maintain Earthquake Sensor Network Connectivity After Loss of Infrastructure}
\author{\IEEEauthorblockN{Gregory R. Steinbrecher}
\IEEEauthorblockA{Department of Electrical Engineering\\
Massachusetts Institute of Technology\\
Cambridge, USA}}

\maketitle

\begin{abstract}
We present the design of a cross-layer protocol to maintain connectivity in an earthquake monitoring and early warning sensor network in the absence of communications infrastructure. Such systems, by design, warn of events that severely damage or destroy communications infrastructure. However, the data they provide is of critical importance to emergency and rescue decision making in the immediate aftermath of such events, as is continued early warning of aftershocks, tsunamis, or other subsequent dangers. Utilizing a beyond line-of-sight (BLOS) HF physical layer, we propose an adaptable cross-layer network design that meets these specialized requirements. We are able to provide ultra high connectivity (UHC) early warning on strict time deadlines under worst-case channel conditions along with providing sufficient capacity for continued seismic data collection from a 1000 sensor network. 
\end{abstract}

\section{Introduction}\label{sec:introduction}
We examine the common problem in sensor network design of mismatched reliability constraints in a network where a small fraction of traffic \emph{must} meet strict delivery guarantees, while the vast majority is subject to much looser requirements. In this paper, we present a case study that demonstrates one possible resolution of this disparity. 

Earthquake monitoring and early warning sensor networks provide an ideal framework to study this problem. While the vast majority of such sensor networks' capcity is dedicated to non-emergency data collection, early warning and low-latency seismic information is of the utmost importance in facilitating rapid response to earthquakes. Decisions about emergency resource allocation rely on knowledge of magnitude and location of the earthquake. As a result, faster seismic data collection can correspond directly to lives saved. 

Projects such as NASA's AIST project and Japan's GEONET have demonstrated the capability to collect such seismic data in real-time by utilizing GPS position data\cite{bock2011}. However, such systems typically rely on local telecommunications infrastructure such as mobile networks or physical connection to landlines. Unfortunately, this is precisely the infrastructure that is unavailable in the wake of the disasters they are designed to detect. 

As an extreme example, the March 2011 earthquake in Japan severely damaged every level of the telecommunications infrastructure. 2 of 3 core fibers, 1.9 million fixed-line links, and 29,000 mobile base stations were damaged. Core fibers were not repaired for two days; other services took months to be fully repaired. Even the planned ``stopgap'' capacity to be implemented in the future has a timescale of hours to be operational\cite{sugino2012}. In order for earthquake sensor networks to be most effective in mitigating the consequences of a disaster, maintaining connectivity \emph{directly after} the event is key.

We present a robust network design that is capable of supporting high throughput non-emergency data collection from a large number of nodes while still guaranteeing low latency emergency notification with ultra-high confidence. Importantly, this is accomplished in the absence of any telecommunications infrastructure by utilizing relatively low cost beyond line-of-sight (BLOS) HF communication via the ionospheric waveguide. As such, our network is tolerant of the extremely high bit error rate (BER) and long time of flight (ToF) conditions necessary to achieve reasonable throughput even in poor ionospheric conditions. Each stage of the design process is evaluated under worst-case conditions to ensure that we meet our ultra-high connectivity guarantee. 

\section{Problem Model}\label{sec:problem-model}
In order to perform this case study, it was necessary to make selections of specific values for model parameters. In this section, we present the chosen requirements, sensor capabilities and the model for the BLOS HF physical layer. However, over-specification of the model could easily have led to exploitation of features that invalidates the adaptability of our design process. As such, we try to parameterize as few features as possible leaving out, for example, assumptions about specific geographic features, node placement, or earthquake location and magnitude.  

\subsection{System Requirements}\label{subsec:network-requirements}

There are two key types of traffic our network aims to support: non-emergency seismic data and emergency early warnings. General seismic data is taken to consist of 10 kilobit reports generated hourly by each sensor in the network. When a sensor detects an emergency, it generates a 500 bit emergency report that must be delivered on a strict deadline. 

Due to the limitations of the HF channel, we cannot cost-effectively (alternatively, spectrum-efficiently) meet time deadlines that match those typically seen in unstressed modern communications infrastructure. Instead, we selected a more modest time deadline of 15 seconds for emergency notification; this includes collection of an emergency report (assumed to be 500 bits long) from a subset of sensors reporting the emergency deemed by the server to be of the most importance or to likely contain a diversity of information. 

\subsection{Sensor Network Hardware}

Current seismic monitoring networks such as the California Real Time GPS Network (CRTN) and Japan's GEONET typically have on the order of five hundred to one thousand sensors\cite{Bock2011a}. For numeric simplicity we aimed to support a maximum network size of 1000 sensors.

To enable time division multiplexing, we assumed each sensor has the capability to maintain low-drift time information, even in the event of GPS signal degradation or loss (e.g. in particularly bad weather). As we are operating on a HF channel, our timescale of interest is microseconds. Specifically, our shortest packet duration will be about 400$\mu s$ (see Section \ref{subsec:packetization}). Assuming GPS signal deprivation for a day, requiring a shift of less than ${1 \over 4}$ of a packet corresponds to $10^{-4}$ s/day, well within reason on modern hardware. 

Given this time information, and utilizing the fact that these sensors are designed to be immobile, we can establish and maintain a record of the ToF delay between sensor and server by having each sensor send it's current time to the server and then returning the $\Delta t$. This only needs to be done once, at time of network initialization (though we also provide capacity for live calibration). When a sensor wishes to communicate to the server at a specific time, it can compensate for the delay such that the signal arrives at the server at the correct time \emph{according to the server's clock}. Assuming channel reciprocity (demonstrated for this physical layer in \cite{zhou2010}), this information can also be calibrated to regular server transmissions. 

Finally, we make the assumption that sensors have the ability to independently discriminate their own importance in the event of an earthquake. This means that, on average, only 16 sensors try to report an emergency when one occurs. This may be implemented in several ways. Purely physically, there is some degree of self-selection merely due to the velocity of seismic waves; sensors spread across an area as large as California or Japan will not all register in the same second. Additionally, some local high speed LOS communications could be implemented between sensors, perhaps sectioning them into classes that discuss emergency detection before reporting via the HF channel. However, we do not speculate here on specific implementation details as it seems highly dependent on deployment conditions.

\subsection{Physical Layer}\label{subsec:physical-layer}

We selected HF radio as our physical layer for two reasons: While we could have implemented a meshed LOS network, it would be just as vulnerable to local degradation as existing infrastructure. Having settled on BLOS communication, cost effectiveness drove us to HF radio. The alternatives, dedicated SATCOM communication and aerial LOS relays, are orders of magnitude more expensive to maintain consistently. 

However, HF propagation in the ionospheric waveguide is hardly an ideal channel. It is noisy, unpredictable, and low capacity compared to typical VHF or UHF LOS channels. We assume the system we are designing is to be installed in an environment similar to California or Japan. Specifically, this means that our system will be located at mid-latitude (as opposed to high latitudes where the ionosphere is significantly less predictable \cite{cannon2000}) and that our sensors will be located no more than 1000 kilometers from the server (A centrally located server in California or Japan easily meets this requirement).

We assume that the base station has a much more powerful transmitter and much more sensitive receiver than the sensors. It could, for example, utilize an antenna array to achieve better multipath rejection. Thus, we limit broadcast capability to the server; sensors do not transmit to one another.

We abstracted the channel into three features: A standard noisy bitstream, the presence of rapid, coherent fading, and very slow (time constant of hours) fading.

\begin{figure*}[t]
\centering
\includegraphics[width=6in]{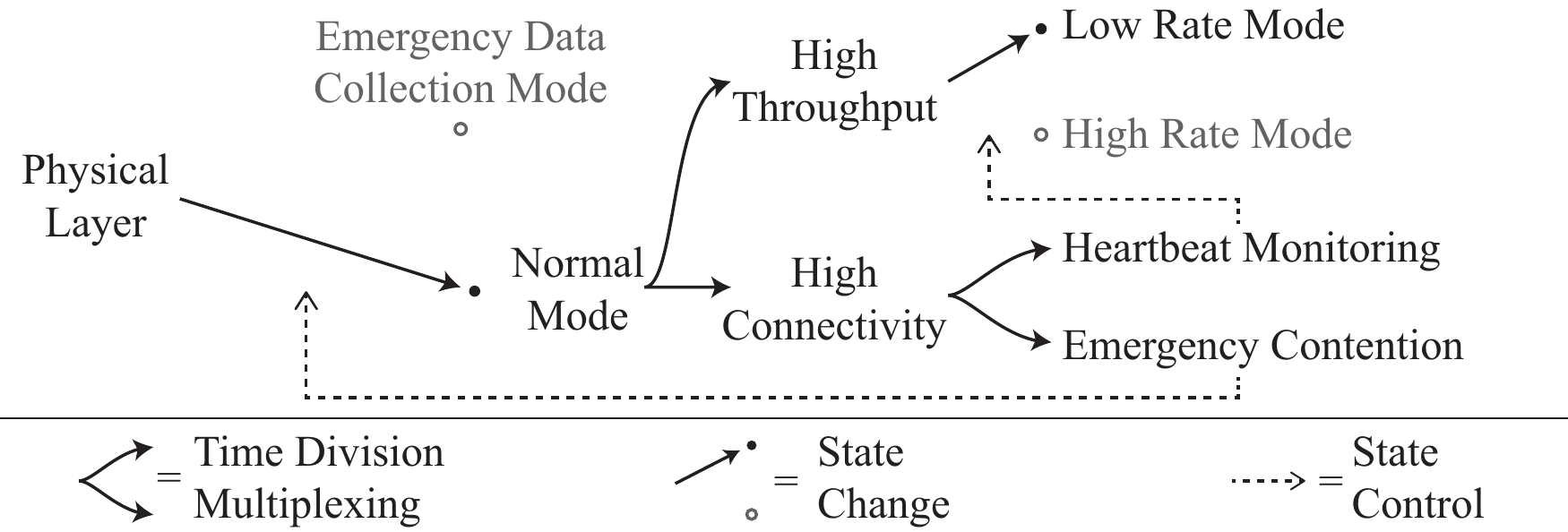}
\caption{System diagram detailing service divisions. Of note is the design decision to allocate 100\% of system capacity to emergency data collection once an emergency is reported. Gray indicates a state that is not currently selected.}
\label{fig:system}
\end{figure*}

\subsubsection{Noisy Bitstream} We assume that we have a radio capable of communication at a bit error rate (BER) of $10^{-2}$ after forward error correction. Depending on the slowly varying channel state, it achieves this BER at 4kbps or 40kbps. Switching between the two modes of operation is assumed to occur in less than 500$\mu s$. Note that this is intended as a loose bound; switching in the digital domain can occur much faster.  

Additionally, signal propagation in the channel is assumed to have a long time ToF. We take this to be relatively constant between fixed points and provide capacity for recalibration to compensate for drift. However, we assume that the ToF can be between 0ms and 20ms for any given sensor. 

Radio links via the ionospheric channel with similar characteristics (datarate, BER, and propagation time) have been experimentally demonstrated\cite{Chamberlain2003}\cite{Herrick1996}. 

\subsubsection{Rapid Fading} Experimental characterizations of high latitude ionospheric HF channels have typically recorded Doppler spreads on the order of a few Hz, with spikes up to 50 or even 100 Hz\cite{cannon2000}\cite{Warrington2009}. Both Japan and California sit closer to mid-latitude, where the channel is expected to be somewhat better behaved. Including the effects of coherent fading in a stochastic analysis is extremely non-trivial. Instead, we consider these effects from the perspective of coherence time. If we assume a typical value of 10Hz, this gives a coherence time ($\tau_{coh}$) of 50ms with the typical convention \cite{Gallager2008} that $\tau_{coh}={1 \over 2D}$. 

\subsubsection{Slow Fading} It is well established that the ionospheric channel varies greatly between night and day\cite{Warrington2009}. Thus, we assume that the channel slowly varies such that, for hours at a time we can achieve 40kpbs operation at the design BER of $10^{-2}$ while for other stretches of hours at a time we can only achieve 4kbps at the same BER. 

\section{System Overview}\label{sec:system-overview}

The main goal of this design is to support two services with disparate reliability and throughput requirements. To cope with this, we introduce service divisions in our network in two ways: time division multiplexing and network state transitions. These divisions are at the core of our design. While the work was unknown when this design was first conceived, the principles introduced in \cite{Chapin2010} helped greatly in clarifying our work. 

At the most basic level, the network can either be in ``Normal Operation'' mode or ``Emergency Data Collection'' mode. Transition from the former to the latter is governed by a contention system: When sensors detect an emergency, they participate in a lottery to notify the server. Once the server detects such a notification, it triggers a switch to emergency mode. Transition back to normal operation occurs only once the server (or decision maker controlling the server) is satisfied sufficient information has been collected. 

As routine data collection is a separate service from emergency reporting, and we cannot support simultaneous collection from all 1000 sensors, we must ensure that sensors maintain their proper configuration. If a sensor is misconfigured (e.g. the ToF has drifted), it may detect an emergency but be unable to communicate that to the server. To ensure reliability of the network, we provide capacity for a second type of critical information: network health monitoring. Following \cite{Chapin2010}, we refer to this as the ``Heartbeat Service.'' 

In the intersts of clarity, we have included a system diagram in Figure \ref{fig:system}.

\section{System Design and Analysis}\label{sec:system-design-and-analysis}

\subsection{Packetization}\label{subsec:packetization}
The high BER property of our channel model means that communication cannot be reasonably conducted via IP packets. Ignoring coherent fading effects and adopting a simple Bernoulli process model for errors, the probability that a packet of length $L_p$ is transmitted without error is $(1-BER)^{L_p}$. Plugging in the BER of $10^{-2}$, a typical bodyless IP packet of 160 bits has a probability of successful transmission of only 20\%. Instead, we implement a shorter packet design that allows us to exploit unique features of a sensor network. 

Given a network of $N$ sensors, we need an address length $A=\lceil\log_2(N)\rceil$ to be able to identify specific sensors. To detect errors, we append a cyclic redundancy check (CRC) to each packet. Calling the number of ``payload'' bits or information-carrying bits per packet $I_p$, our packet CRC needs to be $C_p=\lceil\log_2(L_p)+1\rceil$ bits long in order to detect all errors that have either an odd number of, or less than four, bit flips\cite{Bertsekas1992}.

The expected value of successful packets per packet transmission is $(1-BER)^{L_p}$. Thus, the expected throughput is
\begin{equation}\label{eq:packet-rate}
{R \over L_p} (L_p - C_p - A) (1-BER)^{L_p}
\end{equation}
where R is the datarate in bits/sec. 

\begin{figure}[tb]
\centering
\includegraphics[width=3.5in]{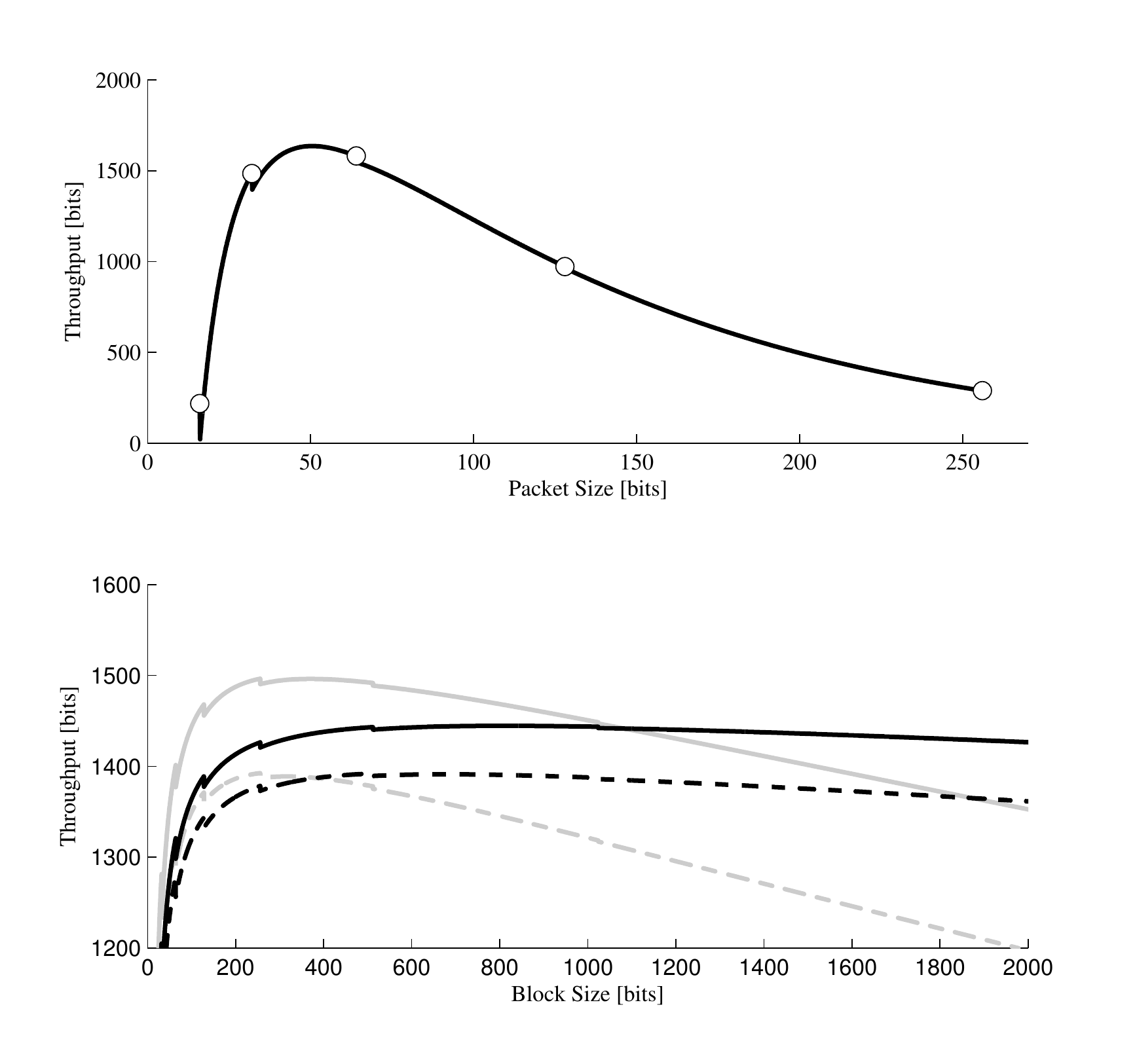}
\caption{(a) Throughput vs. packet size at $BER=10^{-2}$ under a Bernoulli error assumption. 
Circular markers correspond to packet sizes that are a power of 2.
(b) Throughput vs. block size under a Bernoulli error assumption. Black denotes a 32 bit packet, gray a 64 bit packet. Solid lines correspond to $BER=10^{-2}$, dashed to $BER=1.1 \times 10^{-2}$. The discontinuities in both graphs are due to the ceiling function in the calculation of minimum CRC length. }
\label{fig:packsize}
\end{figure}

Plotting this equation (Figure \ref{fig:packsize}a) makes it clear that if we restrict ourselves to packet lengths that are powers of 2, 64 bit packets have the highest expected throughput, though 32 bit packets are only about 5.3\% lower. However, 32 bit packets have several advantages, especially taking into account coherent fading. At the slow transmission rate, a 64-bit packet takes $15.6ms$ to transmit, corresponding to a maximum Doppler spread of 32 Hz, well below the occasional peaks measured in, for example, \cite{cannon2000}. As such, we should expect a reasonable number of extra packets lost to coherent fading with the selection of 64 bit packets over 32 bit packets. 

Additionally, we must take into account the problem of transmitting an entire 10kbit hourly report, not just single packets. We could proceed directly by breaking the 10kbit report into packets and transmitting them one-by-one. However, in the event of undetected packet error, we would have to retransmit the entire 10kbit report. Instead, we first break each report into chunks, each with their own CRC, and then append a final CRC to the report as a whole. 

In order to calculate the throughput of this system, we need to know the probability of an undetected error, $P_u$. As mentioned earlier, a properly selected CRC of length $\lceil\log_2(L_p)+1\rceil$ cannot detect even numbers of bit-flips of 4 or more\cite{koopman2004}. Thus, the probability of an undetected error in a packet is
\begin{equation}
P_u=\sum_{k=2}^{L_p/2} \binom{L_p}{2k} (BER)^{2k} (1-BER)^{L_p-2k}.
\end{equation}

Our expected number of successful blocks per block transmission is $(1-P_u)^{L_b / I_p}$ where the ratio in the exponent denotes the number of successful packets required to transmit one block. The information per block, $I_b$, is equal to $L_b$ minus the length of the necessary CRC, $C_p=\lceil\log_2(L_b)+1\rceil$. 

Bringing this all together, the expected long term throughput is
\begin{equation}
I_b {R \over L_p} {I_p(1-BER)^{L_p} \over L_B} \left( 1-P_u \right)^{L_b / I_p}. 
\end{equation}

We plot this in Figure \ref{fig:packsize}b. Note that a slight bump in BER causes the throughput of the 64 bit packets to drop proportionally twice as much as the 32 bit packets (4.1\% vs. 1.8\%). This suggests their improved performance comes at the cost of decreased robustness to fluctuating channel conditions; an unacceptable trade off in this system. This, in combination with the previous consideration of coherent fading, led to our selection of 32 bit packets. 

We select a 512 bit block over a 1024 bit block for better efficiency in transmitting the 10240 bit hourly report. With 512-bit blocks, we use 43\% of the final (21st) block; with 1024-bit blocks, we use only 12\% of the final (10th) block. As we send the entire block in both cases, the 512-bit block is more efficient. The shorter block also helps to decrease the number of retransmissions in the event of a failed block.

\subsection{Server Scheduled MAC}\label{subsec:server-scheduled-mac}

Because we are interested in earthquake early warning on the timescale of seconds, we divide time into one second frames. At a datarate of 4 kbits/sec and a packet size of 32 bits, this corresponds to 128 packet slots per frame. 

We dedicate 80 packets per frame to transmission of general seismic data collection from sensors to the server, distributed equally among 5 sensors per frame. Splitting the use of the channel among multiple users in this way combats the efficiency loss of MAC scheduling packets being corrupted or lost. 

Which packet slots each sensor (numbered 1-5) broadcast on are constant; the server simply assigns sensors to slots 1 through 5-this fully defines which 16 slots they are to broadcast on. The server performs this scheduling by broadcasting 5 bodyless packets in order with the first packet denoting the user of slot 1 and so on. It repeats these twice to increase the probability that the packet arrives without error at the intended sensor. 

There is a slight possibility of collision here: if a sensor receives its MAC scheduling packet correctly, and another receives one with precisely the four bit flips corresponding to its difference from the intended address, the two stations would try to transmit over one another that frame, resulting in 16 packets lost (and subsequently all NAKed). Combining the tiny probability of this with the fact that the effect is erased the next frame means it does not meaningfully affect the efficiency of the channel. 

This scheduling allows ACK/NAK replies to be handled with a small number of bits as well. Each user sends 16 packets in a given frame, which is precisely the number of payload bits per packet. The server sends 5 ACK/NAK packets (one for each sensor that transmitted), with the corresponding bit representing either NAK (0) or ACK (1) for each of the received packets. We repeat these three times to decrease the probability that a properly received packet is retransmitted or a NAKed packet is misunderstood to have been properly received. 

\subsection{Control and Heartbeat Service}\label{subsec:control-and-heartbeat-service}

Despite the built in repetition noted above, ACK/NAK replies will be misunderstood occasionally. To combat this, we include the capacity for the server to request that a sensor retransmit a specific packet during one of the frame's ``reply'' slots. Requesting such a retransmission is one use for the control packets included in the above frame design; we outline several other uses here:

\begin{itemize}

\item ToF Calibration: If the server recognizes that a sensor's transmissions are impinging on neighboring time slots, it sends the $\Delta t$ to the sensor to recalibrate it

\item General Health Check: Sends a random number to a sensor and ensures that the sensor echoes this number back properly

\item Channel Rate Check: Ask a sensor to attempt communication at high rate, to check if the channel can support it (see Section \ref{subsec:high-rate-operation})

\item Software Updates: For instance, if the operator wishes to redefine the meaning of different packet slots to improve performance in a given environment.

\item Notification that emergency mode has begun

\end{itemize}

As we are only routinely communicating with five sensors per frame, information about the proper operation and configuration of all the other sensors quickly grows stale. If, for instance, the channel has shifted such that a sensor's ToF data is off, this may impair its ability to trigger an early warning. As such, we build in capacity for a ``heartbeat service'' that is used to continuously verify the health and configuration of sensors. 

Due to the limited number of packets per frame, we have included the capability for an arbitrary set control packets multiplexed with the heartbeat packets. This does not violate our goal of separation of services; because there is a reply associated with each control packet, we can use the control packet to verify the proper configuration of the sensor.  

To implement the heartbeat, we send all network control commands encrypted with a sensor's public key. The sensor then decrypts this, verifying that it received the command and processed it correctly, re-encrypts it with the server's public key, and send it back to the server in the appropriate reply slot. 

\subsection{Emergency Mode}\label{subsec:emergency-mode}

As noted previously, when sensors detect an emergency, they enter a contention lottery. When attempting to contend, each sensor randomly selects one of the 16 available slots with uniform probability and attempts to transmit a header-only packet with its address. If we have 16 sensors competing, and there are $n$ slots, the probability that any single slot has exactly one packet broadcast is $\frac{16}{n} (1-{1 \over n})^{15}$. As there are n slots, the expected value of successful slots is $16(1-{1 \over n})^{15}$. 

Given that this is an emergency system, we would like to guarantee a three-9 (99.9\%) worst-case threshold for the time from detection to warning. With 16 slots and 16 contenders, we have a 0.047\% chance of no successful packets. Any fewer contention slots would drop below this threshold. However, in the event that very few sensors are trying to contend, the probability of contention success goes down significantly. To see this, consider only two sensors contending; there is a ${1 \over 16}$ probability that they collide, well below our three-9 threshold. Thus, in the interests of worst-case analysis we assume it takes three seconds for the system to recognize an earthquake and enter the emergency mode. 

The server must then determine which sensors have data to report. We do this by holding two 16 slot (32 half packet slot) contention periods; this ensures that with very high probability (as there are around 16 secondary stations attempting to transmit) we determine all the stations with emergency information. 

We then schedule the winners, starting with any who triggered the initial emergency system, to send in their 500 bit emergency reports. We use an identical framing system as the standard one with a 10 bit CRC on the 500 bit emergency report. We do not batch users; whenever one finishes, we ask another for its report. Once all emergency reports have been collected, we run a final polling period to make sure we did not miss any in the initial contention period. 

The 500 bit emergency reports, with a 10 bit CRC, fit into 32 packets. At the typical BER, this takes 44.1 attempted packet transmissions. Allowing for truly poor conditions, even if it takes 60 packets to collect the error report, five will be completed every four seconds, meaning a maximum of just over 12 seconds to collect all 16 of the relevant emergency alerts, putting us 3 seconds under our design goal of 15 seconds. 

As a summary of the previous several sections, here is the outline of a single frame in 4kbps operation:

\subsubsection{Server Broadcast}
\begin{itemize}
\item 1 Packet of all zeros, indicating channel is in low rate mode
\item 4 Heartbeat/Control Packets
\item Interleaved:

\begin{itemize}
\item 10 MAC Scheduling Half-Packets (5 distinct packets, each sent twice)
\item 15 ACK/NAK Packets for previous frame (5 distinct packets)
\end{itemize}

\item Let channel settle for length of 3 packets
\end{itemize}

\subsubsection{Sensor Transmission}

\begin{itemize}
\item Interleaved from all transmitting sensors:

\begin{itemize}
\item 80 General Seismic Data Packets
\item 4 Heartbeat/Control Replies
\end{itemize}

\item 16 Emergency contention half-packet slots
\item Let channel settle for length of 3 packets
\end{itemize}

\subsection{High Rate Operation}\label{subsec:high-rate-operation}

In order to maintain continuity with the 4kbps scheme, we again work with 1 second blocks. However, as the data rate is ten times faster, we now have 1280 packets to work with. The following frame outline contains the salient details of high speed operation:

\subsubsection{Server Broadcast}
\begin{itemize}
\item 1 Packet of all ones, indicating channel is in high rate mode (duration of 10 packets)
\item 2 Heartbeat Packets (duration of 20 packets)
\item Wait 1 packet duration while all radios switch to 40kpbs operation
\item Interleaved Transmissions:

\begin{itemize}
\item 10 MAC Scheduling Half-Packets (5 distinct packets, each sent twice)
\item 70 Control Packets
\end{itemize}

\item Let channel settle for length of 26 packets
\end{itemize}

\subsubsection{Sensor Transmission}

\begin{itemize}
\item Interleaved from all transmitting sensors:

\begin{itemize}
\item 800 General seismic data packets
\item 70 Replies to Control Packets
\end{itemize}
\item Wait 1 packet duration while all radios switch to 4kpbs operation
\item 16 Emergency contention half-packet slots in slow operation (duration of 80 packets)
\item Let channel settle for length of 30 packets
\end{itemize}

The most import part of this design is that the emergency contention period lines up with ``slow'' operation. Even if a sensor is configured for the wrong datarate, it will still be able to participate in the emergency contention, which always takes place at 4kbps. The same holds for all heartbeat communications.

As a brief summary of other changes, the channel status is monitored by sending out test packets (in the command slots) that are hashed and then returned. When the percent success drops below a certain level (to be determined by in-field testing), the sensor decides the channel has deteriorated and, at the beginning of the next frame, broadcasts a packet of 0's to indicate that sensors should stay in 4kbps mode. 

For the inverse direction, using the heartbeat/control packets in 4kbps operation, the server can indicate that it would like certain sensors to switch to 40kbps for the next frame. Then, the central and relevant sensors all switch to 40kbps operation and the server sends out test packets. This is checked on the third frame's slots 97-104; the hashed versions are returned at 40kbps. If the server determines the channel is fit, it then switches to 40kbps operation. 

\subsection{Network Capacity Analysis}\label{subsec:network-capacity-analysis}

A user needs to send 672 packets successfully to transmit a 10kbit report. At a bit error rate of $10^{-2}$, the probability a 32 bit packet is transmitted without error is very close to 72.5\%. The probability that a packet is transmitted with an undetectable error is approximately 0.027\%. 

As we also have a CRC-10 code for every 502 bits (essentially every 32 packets), the likelihood of having to retransmit the entire hourly report is close enough to zero that our steady state analysis is sufficient. We do expect to have to retransmit a 512 bit chunk one out of every 114 times (or essentially once every six reports). The probability of detectable error is about 27.5\%, meaning that we expect to have to transmit 927 packets per hourly report. At 4kbits/sec, we transmit 16 packets per second per user for five parallel users. This means that, on average, we spend 11.6 seconds per user per hour. In principle this suggests we could support up to 310 sensors solely in the 4kbit/sec mode. Of course, this would fall apart under stressed conditions and is therefore not acceptable for an emergency system.

However, we can support many more users, over the long term, by utilizing the high-rate mode. For the 40kbit/sec mode, we transmit 160 packets per second per user, so we can reasonbly expect to spend about 1.16 seconds per user per hour, or be able to support up to about 3100 users. 

Assuming that the system spends half its time in one state and half in the other, on average we hit about 1700 users, well exceeding our design goal of 1000. This overhead is good; it allows us to deal with things like unexpected corruption, missed frames, etc. And, given the size of the overhead, deviations to below 1000 are exceedingly unlikely. 

\section{Conclusion}\label{sec:conclusion}
As stated at the outset, this work was intended as a case study in networks, particularly sensor networks, that must handle traffic with disparate reliability constraints. The success of our design is predicated on two main features: the model we selected and the service divisions we implemented. 

Model selection is perhaps the most important step in any design process; it is impossible to adequately capture every aspect of the real world in the language of mathematics. As such, it is important to make parameter selections that are as adaptable as possible. This is doubly true when trying to support UHC services; exploiting a quirk in your model quickly leads to unreliable systems when the real world behaves unexpectedly. 

This leads directly to the implementation of service divisions. We were able to take advantage of less reliable capacity (the slowly varying channel state) to gain throughput for less essential services without sacrificing reliability for those which required strict delivery guarantees. 

Utilizing these two features, we have presented and validated a cross-layer network design that meets the design goals. It supports UHC operation for a small subset of traffic while providing sufficient throughput for the less time sensitive, but much higher density, non-emergency data collection. 

\section*{Acknowledgment}
The author would like to thank Vincent Chan for valuable discussions and ideas. Additionally, Matt Carey provided extensive and excellent feedback. 

\bibliographystyle{IEEEtran}
\bibliography{IEEEabrv,steinbrecher-MILCOM2012}

\begin{thebibliography}{10}
\providecommand{\url}[1]{#1}
\csname url@samestyle\endcsname
\providecommand{\newblock}{\relax}
\providecommand{\bibinfo}[2]{#2}
\providecommand{\BIBentrySTDinterwordspacing}{\spaceskip=0pt\relax}
\providecommand{\BIBentryALTinterwordstretchfactor}{4}
\providecommand{\BIBentryALTinterwordspacing}{\spaceskip=\fontdimen2\font plus
\BIBentryALTinterwordstretchfactor\fontdimen3\font minus
  \fontdimen4\font\relax}
\providecommand{\BIBforeignlanguage}[2]{{%
\expandafter\ifx\csname l@#1\endcsname\relax
\typeout{** WARNING: IEEEtran.bst: No hyphenation pattern has been}%
\typeout{** loaded for the language `#1'. Using the pattern for}%
\typeout{** the default language instead.}%
\else
\language=\csname l@#1\endcsname
\fi
#2}}
\providecommand{\BIBdecl}{\relax}
\BIBdecl

\bibitem{bock2011}
Y.~Bock, ``{The Use of Real-Time GPS and Accelerometer Data for Earthquake
  Early Detection and Rapid Response},'' Earthscope 2011 National Meeting,
  Tech. Rep., 2011.

\bibitem{sugino2012}
I.~Sugino, ``{Disaster Recovery and the R\&D Policy in Japan's
  Telecommunication Networks},'' OFC/NFOEC 2012, pp. 19--65, 2012.

\bibitem{Bock2011a}
Y.~Bock, B.~Crowell, D.~Melgar, M.~Squibb, S.~Kedar, F.~Webb, A.~Moore,
  R.~Clayton, and E.~Yu, ``{Real-Time In Situ Measurements for Earthquake Early
  Warning and Spaceborne Deformation Measurement Mission Support In Situ GPS
  Networks for Monitoring},'' Pasadena, CA, pp. 1--24, 2011.

\bibitem{zhou2010}
C.~Zhou, Z.~Zhao, F.~Deng, B.~Ni, and G.~Chen, ``Midlatitude ionospheric {HF}
  channel reciprocity: Evidence from the ionospheric oblique incidence sounding
  experiments,'' \emph{Radio Science}, vol.~45, no.~6, Dec. 2010.

\bibitem{cannon2000}
P.~S. Cannon, M.~J. Angling, N.~C. Davies, T.~Wilink, V.~Jodalen, B.~Jacobson,
  B.~Lundborg, and M.~Broms, ``Damson {HF} channel characterisation-a review,''
  in \emph{{MILCOM} 2000. 21st Century Military Communications Conference
  Proceedings}, vol.~1, 2000, pp. 59--64.

\bibitem{Chamberlain2003}
M.~Chamberlain and W.~Furman, ``{HF Data Link Protocol Enhancements Based on
  STANAG 4538 and STANAG 4539, Providing Greater Than 10kbps over 3kHz
  Channels},'' in \emph{HF Radio Systems and Techniques}, 2003, pp. 64--68.

\bibitem{Herrick1996}
D.~Herrick, ``{CHESS a new reliable high speed HF radio},'' \emph{Military
  Communications Conference, 1996. MILCOM '96, Conference Proceedings, IEEE},
  vol.~3, pp. 684--690, 1996.

\bibitem{Warrington2009}
E.~Warrington, A.~Bourdillon, E.~Benito, and C.~Bianchi, ``{Aspects of HF radio
  propagation},'' \emph{Annals of Geophysics}, vol.~52, no. August, pp.
  301--321, 2009.

\bibitem{Gallager2008}
R.~G. Gallager, \emph{{Principles of Digital Communications}}.\hskip 1em plus
  0.5em minus 0.4em\relax Cambridge University Press, 2008.

\bibitem{Chapin2010}
J.~M. Chapin and V.~W.~S. Chan, ``{Ultra high connectivity military
  networks},'' in \emph{Military Communications Conference}.\hskip 1em plus
  0.5em minus 0.4em\relax IEEE, 2010, pp. 1011--1018.

\bibitem{Bertsekas1992}
D.~P. Bertsekas and R.~G. Gallager, \emph{{Data Networks}}, 2nd~ed.\hskip 1em
  plus 0.5em minus 0.4em\relax Prentice Hall, 1992.

\bibitem{koopman2004}
P.~Koopman, ``{Cyclic redundancy code (CRC) polynomial selection for embedded
  networks},'' \emph{Systems and Networks, 2004}, pp. 145--154, 2004.

\end{thebibliography}
\end{document}